\documentclass[fleqn,usenatbib]{mnras}

\usepackage{newtxtext,newtxmath}

\usepackage[T1]{fontenc}

\DeclareRobustCommand{\VAN}[3]{#2}
\let\VANthebibliography\thebibliography
\def\thebibliography{\DeclareRobustCommand{\VAN}[3]{##3}\VANthebibliography}

\usepackage{graphicx}	
\usepackage{amsmath}	
\usepackage{bm}
\usepackage{xcolor}     
\definecolor{textRed}{rgb}{0.0, 0.00, 0.00}
\newcommand{\textRed}[1]{\textcolor{textRed}{#1}}
\definecolor{textGreen}{rgb}{0.00, 0.0, 0.00}
\newcommand{\textGreen}[1]{\textcolor{textGreen}{#1}}

\title[A new marked correlation function scheme]{\textRed{A new test of gravity - I: Introduction to the method}}

\author[J. Armijo et al.]{
Joaquin Armijo,$^{1,2}$\thanks{E-mail: joaquin.a.armijo-torres@durham.ac.uk}
Carlton M. Baugh,$^{2}$
Peder Norberg$^{2,3}$
Nelson D. Padilla$^{4}$
\\
$^{1}$Center for Data-Driven Discovery, Kavli IPMU (WPI), UTIAS, The University of Tokyo, Chiba 277-8583, Japan\\
$^{2}$Institute for Computational Cosmology, Department of Physics, Durham University, South Road, Durham, DH1 3LE, UK\\
$^{3}$Centre for Extragalactic Astronomy, Department of Physics, Durham University, South Road, Durham DH1 3LE, UK\\
$^{4}$Instituto de Astromía Teórica y Experimental (IATE), CONICET
Universidad Nacional de Córdoba, Laprida 854, X5000BGR, Córdoba, Argentina
}

\date{Accepted XXX. Received YYY; in original form ZZZ}

\pubyear{2022}

\begin{document}
\label{firstpage}
\pagerange{\pageref{firstpage}--\pageref{lastpage}}
\maketitle

\begin{abstract}
We introduce a new scheme based on the marked correlation function to probe gravity using the large-scale structure of the Universe. We illustrate our approach by applying it to simulations of the metric-variation $f(R)$ modified gravity theory and general relativity (GR). 
The modifications to the equations in $f(R)$ gravity lead to changes in the environment of large-scale structures that could, in principle, be used to distinguish this model from GR. Applying the Monte Carlo Markov Chain algorithm, we use the observed number density and two-point clustering to fix the halo occupation distribution (HOD) model parameters and build mock galaxy catalogues from both simulations. To generate a mark for galaxies when computing the marked correlation function we estimate the local density using a Voronoi tessellation. 
Our approach allows us to isolate the contribution to the uncertainty in the predicted marked correlation function that arises from the range of viable HOD model parameters, in addition to the sample variance error for a single set of HOD parameters. This is critical for assessing the discriminatory power of the method. In a companion paper, we apply our new scheme to a current large-scale structure survey.  
\end{abstract}




\section{Introduction}\label{sec:introduction}

The cosmological constant $\Lambda$ was initially introduced by Einstein to produce a stationary universe solution to his field equations (for a historical review see \citealt{ORaifeartaigh2017}). However, the nature of this component, now thought to be responsible for the accelerated cosmic expansion, remains unknown 
and may point to the need to change the theory of gravity \citep{Jain2013,Heymans2018,Baker2021b}. The evolution of the Universe after the Big Bang is imprinted on the large-scale structure, also called the cosmic web, through the interplay between gravity and the expansion rate. 
This means that the large-scale structure of the Universe not only contains information about the cosmological model but also about the nature of gravity on cosmological scales. Moreover, many models that modify the theory of gravity from general relativity replicate the accelerated expansion without invoking a cosmological constant \citep{Clifton2012,Joyce2015,Koyama2016,FADE2022,Martinelli2021}. For such models, new degrees of freedom are introduced, which must be coupled to matter, altering the formation of structure over time compared to GR. Alternative models of gravity will inevitably modify structure formation in a manner that depends on the environment.

Observational constraints on gravity, such as from the dynamics of the solar system, and the more recent detection of binary neutron star mergers, have led to several classes of modified gravity (MG) models that were until recently under consideration now being ruled out \citep{LombriserTaylor2016,Creminelli&Filippo2017,Ezquiaga2017,Baker2021a}. Such theories of gravity modify the propagation velocity of  gravitational waves detected in a vacuum, which is inconsistent with the current detections of ``multi-messenger'' events. However, many models of modified gravity are still viable as they are allowed by local tests of gravity in the solar system and on galactic scales. Such models are continually being tested and further constrained, and include chameleon theories, for example $f(R)$ gravity \citep{Applebly2007,DeFelice2010} and Brans-Dicke type theories including the Dvali-Gabadadze-Porrati (DGP) model \citep{Dvali2000}. These modified gravity models are important as they can be used to test GR and the equivalence principle on cosmological scales. In the last decade, N-body simulations of modified gravity have been developed to study MG in the large-scale Universe, allowing new probes of gravity to be explored  \citep{Li2011,Winther2014,Arnold2019b}

 To further constrain MG models, we need probes that are sensitive to changes in the environment of large-scale structures brought about by the changes to gravity, compared with GR. These impacts are seen in phenomena such as weak lensing \citep{Kilbinger2015,Ruth2022}, redshift space distortions \citep{Jennings:2012,Ruan2022}, and the marked correlation function statistic \citep{White:2016,Aviles2020}. Here, we focus on the latter, which is a relatively new statistical tool that contains information beyond the traditional galaxy-galaxy correlation function, whilst still being a second moment quantity. The marked correlation function has been used to study the connection between properties of galaxies and their environment, such as luminosity and environmental density, and halo mass \citep{ShethTormen2004,Percival2004,Wechsler2006}. 

Our main aim here is to introduce a new cosmological probe of gravity, a marked correlation function in which the mark depends on density. To meet this aim we have developed a pipeline to make realisations of mock galaxy catalogues from N-body simulations, using a simple halo model approach. A key feature of our analysis is an assessment of  
the uncertainty in the model predictions due to the range of halo models that give acceptable fits to the measured two-point correlation function and the number density of the tracers; this uncertainty is often ignored in the literature and could result in an overly optimistic view of the performance of any diagnostic that depends on clustering \citep{Armijo2018,Hernandez-Aguayo2019,Valogiannis2018,Satpathy2019}. Here, we introduce the new methodology, which improves upon the modelling developed in \cite{Armijo2018}. We apply the approach introduced in this paper to current surveys in a companion study (Armijo et al. 2023, hereafter Paper II); here we show results for one of the samples considered in Paper II to illustrate the method and leave the discussion of the details of how the mock catalogues are constructed to that paper. 

The outline of this paper is as follows: in $\S$~\ref{sec2} we give an overview of the $f(R)$ theory of gravity, which is the model we use to compare to GR and hence to illustrate our method. The simulations used to understand the modelling of modified gravity are presented in $\S$~\ref{sec3} and the creation of mock galaxy catalogues to replicate the observations is described in $\S$~\ref{sec4}. The calculation of the marked correlation function is presented in $\S$~\ref{sec5}. Finally, we explain the direction in which this work could go in the future and draw our conclusions in $\S$~\ref{sec6}.

\section{The \textit{f(R)} theory of gravity} \label{sec2}
The $f(R)$ theory of gravity \citep{Sotiriou2010} is a viable alternative to general relativity. In the standard $\Lambda$CDM model, the cosmological constant, $\Lambda$, drives the accelerated expansion of the universe at recent times. Instead of invoking $\Lambda$, $f(R)$ gravity models explain the quickening expansion by invoking new physics that arises from the additional degrees of freedom introduced into the equations of motion for gravity (see for example \citealt{Li2007}). 

The $f(R)$ model of gravity can be viewed as an extension of standard GR through the inclusion of a function, $f$, of the Ricci scalar, $R$, in the Einstein-Hilbert action

\begin{equation}
    S = \int d^4x \sqrt{-g} \left( \frac{1}{2\kappa^2}\left[R + f(R)\right] + \mathcal{L}_m \right),\label{chp2:eq1}
\end{equation}
where $k^2 = 8\pi G$, $G$ is Newton's constant, $g$ is the determinant of the metric $g_{\mu\nu}$ and $\mathcal{L}_m$ is the Lagrangian density of matter. The form of the $f(R)$ function can be chosen to mimic the expansion history of the $\Lambda$CDM model which is well constrained by observations of the cosmic microwave background and the large-scale structure in the galaxy distribution. The addition of this extra term in Eqn.~\ref{chp2:eq1} leads to the modification of all the equations of GR, including the Einstein field equations

\begin{equation}
    G_{\mu\nu} + f_RR_{\mu\nu} - g_{\mu\nu} \left[ \frac{1}{2}f - \nabla^2 f_R \right] - \nabla_{\mu}\nabla_{\nu}f = 8\kappa T_{\mu\nu}, \label{chp2:eq2}
\end{equation}
where $\nabla_{\mu}$ is the covariant derivative of the metric tensor, $f_R \equiv {\rm d}f(R)/{\rm d}R$ is the new scalar and dynamical degree of freedom that arises from the introduction of the $f(R)$ term. To solve this new equation and obtain the equations of motion for massive particles, one can take the trace of Eqn.~\ref{chp2:eq2} and solve for the case of a perturbation around the standard Friedmann-Lema\^itre-Robertson-Walker metric. This description of the background evolution of the universe gives two equations of motion. The first is the modified Poisson equation:

\begin{equation}
    \Vec{\nabla}^2 \Phi = \frac{16\pi G}{3}a^2[\rho_m - \bar{\rho}_m] + \frac{1}{6}a^2 \left[ R(f_R) - \bar{R} \right], \label{chp2:eq3}
\end{equation}
and the other is for the new scalar field, $f_R$:

\begin{equation}
    \Vec{\nabla}^2 f_R = -\frac{1}{3}a^2\left[ R(f_R) - \bar{R} + 8\pi G(\rho_m - \bar{\rho}_m) \right], \label{chp2:eq4}
\end{equation}
where $\rho_{\textrm{m}}$ is the matter density field, and an overbar indicates quantities ($\bar{\rho}_m$ and $\bar{R}$) defined as mean values for the background cosmology. As we have now defined the Ricci scalar as a function of $f_R$ in both Eqns~\ref{chp2:eq3} and \ref{chp2:eq4}, we can combine these to obtain 

\begin{equation}
    \Vec{\nabla}^2 \Phi = 4\pi G a^2 \left[ \rho_m - \bar{\rho}_m \right] - \frac{1}{2}\Vec{\nabla}^2 f_R \label{chp2:eq5},
\end{equation}
which is a new equation of motion for massive particles including a term which comes from the new scalar degree of freedom. We can understand this new term as the potential $- 1/2 f_R$ of an extra force, the fifth force, mediated by the scalar field $f_R$, which is sometimes referred to as the scalaron \citep{GANNOUJI2012}.

\subsection{The chameleon mechanism}
The equations of motion of $f(R)$ gravity are different from those in standard gravity, and different predictions may result. Nevertheless, local tests already constrain these predictions with great accuracy on certain scales, such as in the solar system \citep{Guo2014}. This means that modified gravity must include mechanisms to hide the new physics which arises from the extra degree of freedom in Eqn.~\ref{chp2:eq5} on these scales. This feature is referred to as a screening mechanism \citep{Khoury2004}, and is a scale-dependent property of chameleon theories such as $f(R)$ gravity. On scales where the model is expected to behave as standard gravity, such as in the deep Newtonian potential of the Solar system, Eqn.~\ref{chp2:eq4} is dynamically driven to $\left| f_R \right| \rightarrow 0$. In this limit, Eqn.~\ref{chp2:eq5} reduces to the standard Poisson equation and GR is recovered, hence this theory is viable on these scales \citep{HuSawicki2007}. On the other hand, on scales where the Newtonian potential becomes shallower, the term $R - \bar{R}$ in Eqn.~\ref{chp2:eq4} is negligible and Eqn.~\ref{chp2:eq5} reduces to
\begin{equation}
    \Vec{\nabla}^2 \Phi= \frac{16}{3}\pi G a^2 [\rho_m - \bar{\rho}_m],
\end{equation}
which is the same as the standard Poisson equation, but enhanced by a factor $4/3$ when the amplitude of the fifth force is at its maximum and no screening is triggered. An interesting feature of this theory is that to obtain Eqn.~\ref{chp2:eq5} no assumption about the form of the $f(R)$ function is required, which means that this mechanism is independent of the choice of $f(R)$.

\subsection{The Hu \& Sawicki model}
A popular choice for the functional form of $f(R)$ is the one proposed by \cite{HuSawicki2007}
\begin{equation}
    f(R) = -m^2\frac{c_1 \left( \frac{R}{m^2} \right)^n }{c_2 \left( \frac{R}{m^2} \right)^n +1},
    \label{chp2:eq6}
\end{equation}
where $m^2 = 8\pi G\bar{\rho}_{m0}/3$ is called the mass scale, $\bar{\rho}_{m0}$ is the value of the background matter density today, and $n$, $c_1$ and $c_2$ are free parameters of the model. The form of this function is motivated by the aim of ensuring that for high curvature values compared to the mass scale, $m^2$, the term $m^2/R$ goes to zero and $f(R)$ can be expanded as 

\begin{equation}
    f(R) \approx -\frac{c_1}{c_2}m^2 + \frac{c_1}{c_2^2}m^2 \left( \frac{m^2}{R} \right)^n.
    \label{chp2:eq7}
\end{equation}
In the limit $m^2/R \rightarrow 0$, the term $c_1/c_2$ acts as the cosmological constant of this model, and is independent of scale. As we have an explicit form for  $f(R)$ we can set $c_1/c_2 = 6\Omega_{\Lambda,0}/\Omega_{m,0}$, where $\Omega_{m,0}$ is the matter density parameter today, and $\Omega_{\Lambda} = 1 - \Omega_{m}$. With this configuration, the model follows the same expansion history as the $\Lambda$CDM model by construction. Meanwhile, the scalaron field can also be approximated by
\begin{equation}
    f_R \approx -n \frac{c_1}{c_2^2}\left(\frac{m^2}{R}\right)^{n+1},
    \label{chp2:eq8}
\end{equation}
and we can also evaluate the expansion history today, where $R_0 \gg m^2$. \textRed{In this scenario,  the scalaron solution of Eqn. \ref{chp2:eq4} sits in the minimum of the effective potential, then the Ricci scalar can be solved using the background values \citep{Brax2008}} 

\begin{equation}
    \bar{R} \approx 8\pi G \rho - 2\bar{f}(R) = 3m^2\left[ a^{-3} + \frac{2}{3}\frac{c_1}{c_2} \right],
\end{equation}\label{chp2:eq9}
\textRed{which removes the dependence between $R(f_R)$ and the scalaron $f_R$. Then this approximation can be used to solve the term $c_1/c_2^2$ in Eqn \ref{chp2:eq8}}:
\begin{equation}
    \frac{c_1}{c_2^2} = -\frac{1}{n}\left[ 3 \left( 1 + 4\frac{\Omega_{\Lambda}}{\Omega_m} \right) \right]^{n+1} f_{R0},
\end{equation}
which is evaluated with the value of the scalaron today, $f_{R0}$. By fixing these values the model depends on only two free parameters, $n$ and $f_{R0}$. These parameters can be constrained using the large-scale structure at late times. One of the fundamental measurements to obtain these constraints is the power spectrum for a range of models with different values of the scalaron amplitude $\left| f_{R0} \right|$ when fixing $n=1$. 

\section{N-body simulations and mock catalogues.}\label{sec3}

In this section we describe the N-body simulations used (\S~3.1), the halo catalogues extracted from them (\S~3.2), and 
the HOD framework adopted to populate the halos with galaxies (\S~3.3).

\subsection{Simulations of modified gravity}
We use simulations of the cold dark matter cosmology with different gravity flavours, standard GR and modified $f(R)$ gravity from \cite{Arnold:2019}. These calculations use the 2016 Planck cosmological parameters  \citep{Planck:2016}: $h = 0.6774$, $\Omega_{\textrm{m}} = 0.3089$, $\Omega_{\Lambda} = 0.6911$, $\Omega_{\textrm{b}} = 0.0486$, $\sigma_8 = 0.8159$, and $n_{\textrm{s}} = 0.9667$. We use a model of $f(R)$ with amplitude $\left| f_{R0} = 10^{-5} \right|$ denoted as F5 and a model of standard general relativity referred to as GR. These simulations use $2048^3$ collisionless particles in cubic boxes of length $L_{\textrm{box}} = 768\, h^{-1}\, \textrm{Mpc}$ resulting in a particle mass of $M_{\textrm{p}} = 4.9 \times 10^{9} h^{-1}\ \textrm{M}_{\odot}$.  Here we use the simulation output at redshift $z=0.3$. 

\subsection{Haloes and subhaloes}
Haloes are identified using the \textsc{subfind} algorithm \citep{Springel:2001}. In the first step, the friends-of-friends (FoF) percolation scheme is run on the simulation particles in a given snapshot. The minimum number of particles per group retained after the FoF step is set to 20. \textsc{subfind} is then applied to find the local density maxima in the FoF particle groups, and checks to see if these structures are gravitationally bound. Unbound particles are removed from the membership list. The resulting objects are called subhaloes, which correspond to haloes which fell into a more massive structure at an earlier time and are still in the process of merging with it. We use the positions of these haloes and subhaloes to populate the simulation box with central and satellite galaxies, rather than resorting to sampling spherically symmetric NFW profiles which end at the virial radius, as has been used in many previous studies (e.g. \citealt{Cautun2018,Paillas2018,Armijo2018,Hernandez2018}). This choice was made to achieve better agreement between the mock catalogues and the observations, particularly on small scales. 
This saves us the step of creating a halo profile for individual haloes, which would introduce an extra parameter, the concentration, to model the position of satellite galaxies. Taking our approach instead allows the HOD parameters to be constrained more tightly. 

\subsection{HOD galaxy catalogues}
The HOD model \citep{Peacock2000,Berlind2002} is an empirical description of the number of galaxies per halo as a function of halo mass. By using the simulated halo and subhalo catalogues we aim here to recreate the BOSS LOWZ sample  \citep{Dwason:2013} (we consider this sample along with the  CMASS LRG sample of \cite{Reid2016} in Paper II, to which we refer the reader for further details). The HOD prescription gives the number of central and satellite galaxies separately as functions of halo mass \citep{Zheng2007}:

\begin{eqnarray}
    \left< N_{\textrm{cen}} \right> & = & \frac{1}{2}\left[ 1 + \text{erf} \left( \frac{\log M - \log M_{\textrm{ min}}}{\sigma_{\log M}} \right) \right] \label{chp5:eq:HOD_cen} \\
    \left< N_{\textrm{sat}} \right> & = &\left< N_{\textrm{cen}} \right> \left( \frac{M-M_0}{M_1} \right)^{\alpha}. \label{chp5:eq:HOD_sat}
\end{eqnarray}
In Eqn.~\ref{chp5:eq:HOD_cen}, $N_{\textrm{cen}}$ is the mean number of central galaxies as a function of the mass of the halo, $M$, and $M_{\textrm{ min}}$ and $\sigma_{\log M}$ are free parameters. In the case of satellites, Eqn.~\ref{chp5:eq:HOD_sat} is dependent on Eqn.~\ref{chp5:eq:HOD_cen} and $M$, because the satellite population of the halo is linked to whether or not there is a central galaxy. $M_0$, $M_1$, and $\alpha$ are free parameters in Eqn.~\ref{chp5:eq:HOD_sat}. 
In cases where the number of satellites is higher than the subhaloes attached to an individual halo, the subhaloes are recycled as satellite hosts, and could, in principle, host more than one satellite galaxy. However, this effect is at the sub-percent level for the HOD parameters used.

\section{Inferring HOD parameters using the Markov Chain Monte Carlo  method}\label{sec4}

The HOD framework provides a simple and accurate means of describing a galaxy population defined by a set of selection criteria, to allow a reproduction of the large-scale structure measured in a wide field survey for a given space density of tracers. Here, to illustrate our method, we consider the application of the HOD framework to luminous red galaxies (LRGs) from the Sloan Digital Sky Survey (SDSS), which have been used to trace the large-scale structure efficiently over a large volume of the Universe \citep{Eisenstein:2001,Eisenstein2011}, focusing on the LOWZ sample of \cite{Parejko2013}. 
The objective is to build mock catalogues that match the number density and projected clustering \textGreen{measured for} galaxies in the SDSS LRG sample from both the modified gravity and GR simulations. \textRed{Further details of the LOWZ sample are provided in Paper II.} 

Several studies have been performed to construct such mock galaxy catalogues \citep{Parejko2013,Manera2013,Manera2015}. Here, we try to improve on the procedure used by Parejko et al. in several ways: first, we restrict the redshift range of the samples to reduce the variation in the observed number density, allowing this property to be modelled more accurately, and second, we develop a new scheme for fitting the observed number density along with the clustering. We describe our method in the next section. Another method worth mentioning is that presented by \cite{Zhang:2022}, in which the HOD parameters are constrained using high-order clustering statistics. The combination of the two and three-point functions used by Zhang et~al. allows, in principle, tighter constraints to be placed on the HOD parameters than when using the two-point function alone. However, the estimation of the three-point correlation function is significantly more time-consuming than two-point functions \citep{Guo2015} and so this option is not considered further here, as our pipeline involves estimating the clustering for tens of thousands of mock catalogues.

\begin{figure*}
    \centering
    \includegraphics[width=0.99\textwidth]{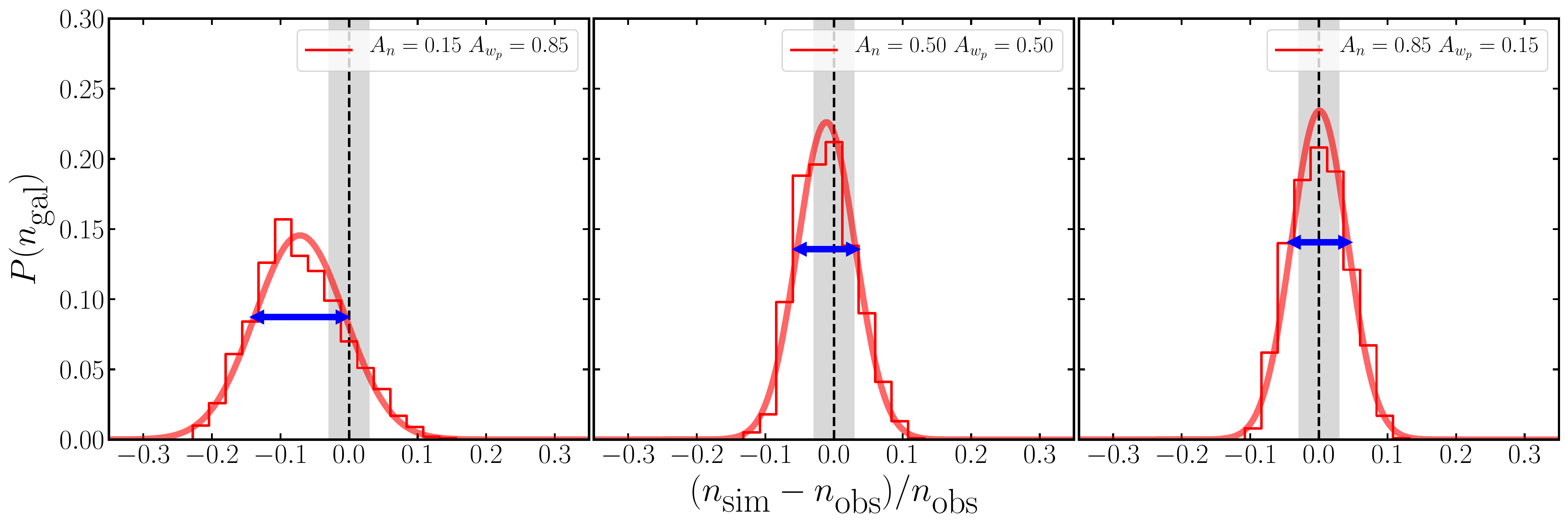}
    \caption{The distribution, \,$P(n_\textrm{gal})$, of the galaxy number density, $n_\textrm{sim}$, recovered for the HOD samples for the different weighting schemes (red histogram): $A_\textrm{n}=0.15$, $A_{w_{\textrm{p}}}=0.85$ (left panel); $A_\textrm{n}=0.50$, $A_{w_{\textrm{p}}}=0.50$ (middle panel) and $A_\textrm{n}=0.85$, $A_{w_{\textrm{p}}}=0.15$ (right panel). We draw over each $P(n_\textrm{gal})$ a Gaussian with the same mean and standard deviation as the distributions (smooth red curve). We have rescaled the $x$-axis to the relative difference between the individual measurements of $n_\textrm{sim}$ and the target $n_\textrm{obs}$ (black dashed line) for the LOWZ sample. We show the uncertainty in the value of $n_\textrm{obs}$ (grey shaded area), which is calculated using jackknife resampling. \textGreen{We mark the 1-$\sigma$ range for the red curves in each panel (blue line ending in arrows)} to highlight the deviation between the $n_\textrm{sim}$ distribution and the target $n_\textrm{obs}$ for the choice of weight used in each panel.}
    \label{fig:n_gal}

\end{figure*}

\begin{figure}
    \centering
    \includegraphics[width=0.45\textwidth]{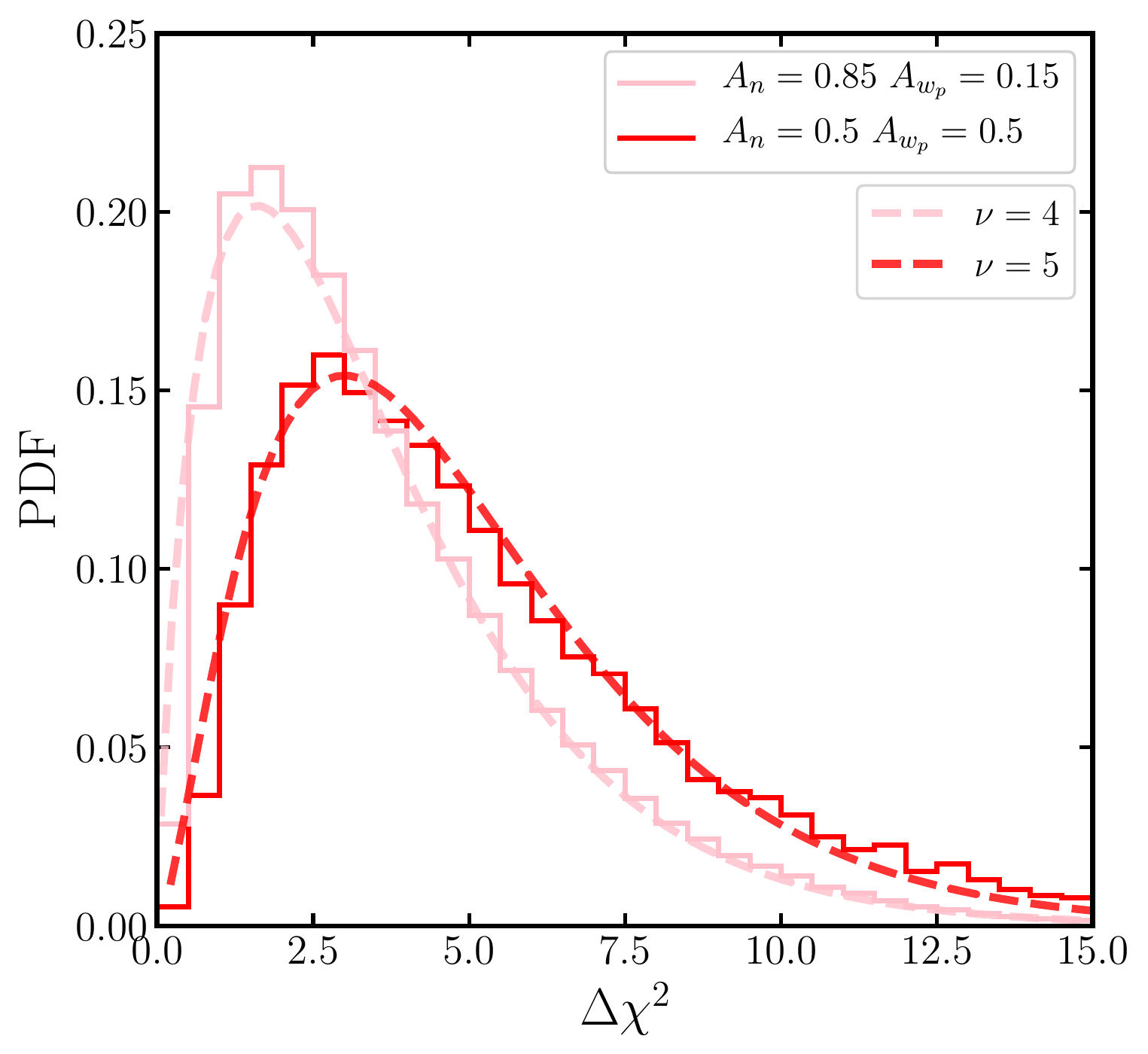}
    \caption{The $\chi^2$ distribution for the MCMC chains of the HOD fits. We show two cases of the weight scheme, which produce similar results for $n_{\rm gal}$ and $w_{\rm p}$ :$A_{\rm n}=0.85$, $A_{\rm w_p}=0.15$ (pink histogram) and $A_{\rm n}=0.50$, $A_{\rm w_p}=0.50$ (red histogram). The smooth curves show the corresponding analytical $\chi^2$ distributions that best represent the data for each case, with $\nu = 4$ (pink dashed line) and $\nu = 5$ (red dashed line).}
    \label{fig:chi2dist}
\end{figure}

\begin{figure*}
    \centering
    \includegraphics[width=\linewidth]{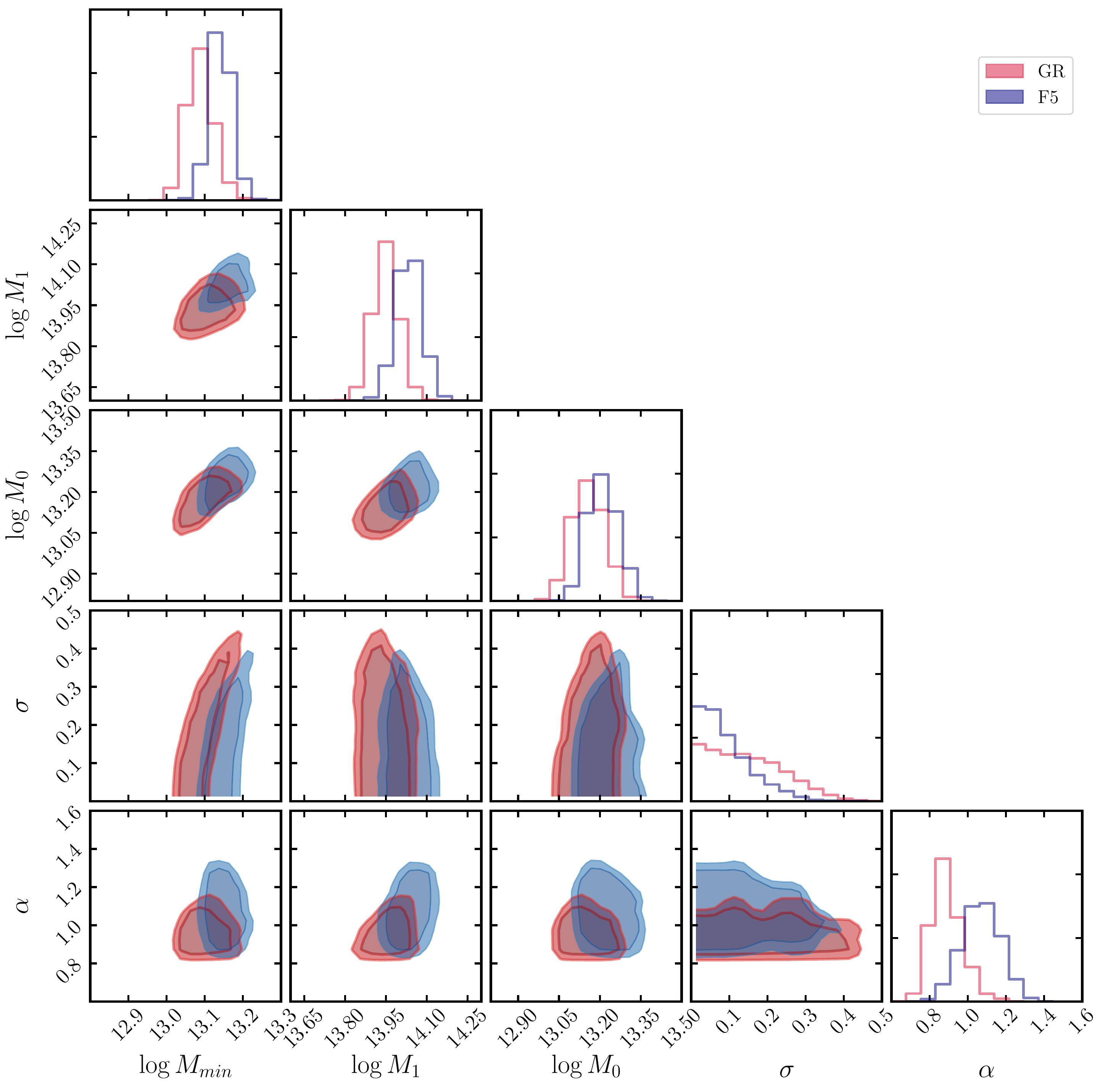}
    \caption{Corner plot showing the MCMC posterior distribution for the HOD model parameters (with the units given in Table~\ref{table:priors}), for the fit to the LOWZ data. We use the MCMC method to fit the HOD model from either the GR (red) or F5 (blue) simulations to the data we want to replicate (in this illustration, LOWZ).%
    \, The diagonal subpanels show the 1-D distribution of the parameters of the posterior distribution, $p(\theta)$, with $\theta$ being the HOD parameters. The off-diagonal subpanels show the 2-D projection of the parameters for all parameter combinations, where the contours are selected using $\Delta\chi^2$, using $1$-$\sigma$ (inner lines) and $2$-$\sigma$ (outer lines), which correspond to $\Delta \chi^2$ of $2.31$ and $6.17$ respectively.}
    \label{fig:MCMC_posterior}
\end{figure*}

\begin{figure}
    \centering
     \includegraphics[width=0.45\textwidth]{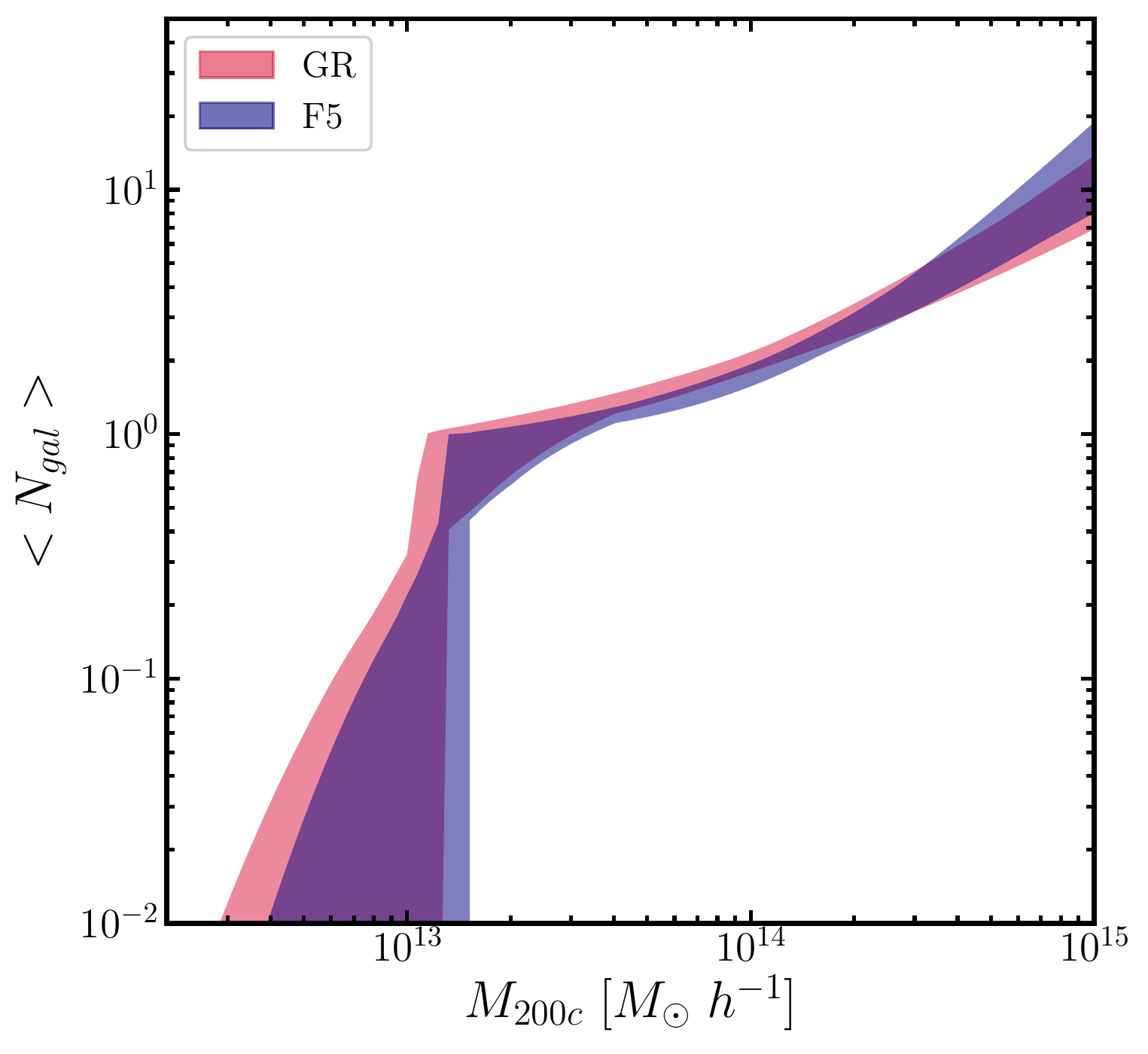}\\
    \caption{The expected number of galaxies in a halo, $ \left< {N} \right> $, as a function of halo mass $M_{200c}$ for all the HOD parameter sets which lie within a $1 \, \sigma$ confidence interval according to the $\chi^2$ distribution. We show the HOD region for both GR (red) and F5 (blue) models, selecting the best 68\% from the $\Delta \chi^2$ distribution with $\nu=5$.}
    \label{fig:HODcurves}
\end{figure}

\begin{table}
    \centering
    \resizebox{\linewidth}{!}{
    \begin{tabular}{|p{3cm}|| p{3cm} |}
    \hline
    \hline
    \multicolumn{2}{|c|}{HOD parameter-space adopted for GR and F5 simulations} \\
    \hline
    \multicolumn{2}{|l |}{$\theta$} \\
    \hline
    $\log (M_{\rm min}\ /\ h^{-1} {\rm \ M_{\odot}} ) $ & $[12.7,14.0]$\\
    $\log (M_{1}\ / h^{-1} {\rm \ M_{\odot}})$ & $[12.7,14.8]$ \\
    $\log (M_{0}\ /\ h^{-1} {\rm \ M_{\odot}} ) $ & $[12.7,14.0]$\\ 
    $\sigma_{\log M}$ & $[0.0,0.6]$\\
    $\alpha$ & $[0.4,1.6]$\\
    \hline
    \hline
    \end{tabular}
    }
    \caption{The uniform priors adopted for the HOD parameter set, $\theta$. Extra conditions are applied to some of the prior distributions, such as requiring that $M_0 > M_{\rm{min}}$ and that $ M_1 > 5\,M_0$ for every set of HOD parameters.}
    \label{table:priors}
\end{table}

We use the Metropolis-Hasting MCMC scheme \citep{Metropolis1953,Hastings1970} to explore the 5-dimensional HOD parameter space, and obtain the best fitting parameters that replicate the number density and clustering of the LOWZ sample, as an example of how we can fit observations using a metric that depends on both quantities. \textRed{We define the log-likelihood for each quantity, proportional to the $\Delta\chi^2$ distribution, where $\chi^2$ is defined by}

\textRed{
\begin{equation}
    \chi_{\mu}^2 = (\mathbf{x} - \bm{\mu})^{\intercal} \mathbf{\Sigma^{-1}} (\mathbf{x} - \bm{\mu}),\label{chp5:eq:chi_2_def}
\end{equation}
}\textRed{where $\mathbf{x}$ is the realization value drawn from the set of parameters, and $\bm{\mu}$ is the observable that we are trying to model. $\mathbf{\Sigma^{-1}}$ is the inverse of the covariance matrix, which includes the uncertainties in the observation of $\bm{\mu}$. The above definition is valid for the two observable quantities we are trying to fit, $n_{\rm gal}$ and $w_{\rm p}$}. As we are trying to fit two quantities that are related \textGreen{at some level}, such as the clustering and number density of the galaxy sample, we need to consider this when defining the $\chi^2$ that we want to measure. \textGreen{Here, w}e define  a new, \textGreen{phenomenological form of}  $\chi^2$ by combining both measurements:
\begin{equation}
    \chi^2 = A_{\rm n}\chi_{\rm n}^2 + A_{w_{\rm p}}\chi_{w_{\rm p}}^2, \label{chp5:eq:chi_2_weighted}
\end{equation}
where $A_{\rm n}$ and $A_{w_{\rm p}}$ are factors that weight the individual $\chi^2$ for the number density, $n$, and the clustering, $w_{\rm p}$, respectively. By adding these quantities, we can fit models to the data using the adopted weights for these two metrics, which in turn can provide a better understanding of the correlation between the clustering and number density, and help us to determine if one is more important than the other when looking for the best fitting HOD parameters. \textGreen{This is a phenomenological, pragmatic solution which we will demonstrate using our mock catalogues. We aim to choose a set of weights that give an unbiased recovery of the number density of galaxies and their clustering. Furthermore, the results should not be sensitive to the precise value of the weights, and, for this reason, we use the same weights for different gravity models. Our definition of $\chi^{2}$ scales with the number of bins in the statistic being probed. Thus, without using weights, the correlation function, which is measured in many bins, would dominate over the number density measurement. We argue below that some weighting is necessary to improve the accuracy with which the number density is recovered. It is important to consider the number density as a constraint; if the mocks for the gravity models had different galaxy densities, this could lead to differences in the marked correlation functions that are not due to gravity.}

We need to determine the weights, $A_{\rm n}$ and $A_{w_{\rm p}}$, and the range of acceptable HOD parameters for each model. We used {\ttfamily emcee} \citep{emcee}, which is a {\ttfamily Python} implementation of the MCMC algorithm that applies the Metropolis-Hasting ensemble sampler. We build the ensemble using 28 walkers each running for $30\,000$ iterations ($10\,000$ for the so-called burn-in or settling down phase and $20\,000$ for the production phase used to estimate the posterior distribution); these choices are motivated by using the autocorrelation time analysis and the Gelman-Rubin (G-R) diagnostic \citep{Gelman1992}. For the autocorrelation time, $\tau_{\rm f}$, we estimate the convergence at $N=50\tau_{\rm f}$ iterations, for all the chains tested. We also calculate the G-R diagnostic for the chains. We provide the parameter space limits applied to the priors, used for searching the HOD parameters, in Table~\ref{table:priors}. To investigate the \textGreen{impact of the} choice of weights, we try three runs with different $\chi^2$ definitions: $A_{\rm n} =0.15$, $A_{w_{\rm p}}=0.85$; $A_{\rm n}=0.85$, $A_{w_{\rm p}}=0.15$ and $A_{\rm n}=0.5$, $A_{w_{\rm p}}=0.5$. These cases are useful to study over what range we can adjust the metrics without introducing biases into the recovered parameter values and statistics. 

 \subsection{The HOD families that reproduce LOWZ results}
The clustering of galaxies is a robust probe of the cosmic large-scale structure and the increasingly accurate measurements that have been made over the past twenty years have played an important role in constraining the basic cosmological parameters  \citep{Percival2001,Cole2005,Sanchez2009,Reid2010,Ross2012a,Alam2015,Icaza2020}. When considering alternatives to GR-$\Lambda$CDM, the predicted two-point correlation function and abundance of galaxies should agree with existing measurements. Hence an approach that is becoming increasingly common in the literature is to choose HOD model parameters such that the abundance and projected two-point correlation function of a variant model look as similar as possible to those in 
GR \citep{Cautun2018,Paillas2018}.

To search for the HOD parameters that give us mock galaxy samples that mimic the number density and clustering of the LOWZ sample, we need to explore how the choice of weights in the goodness of fit metric affects the recovered statistics. For instance, the number density, which is the mean number of galaxies per unit volume, is represented by one number for every HOD sample, ($n_{\textrm{gal}} = N_{\textrm{gal}}/V_{\textrm{box}}$), where the volume of the simulation is $V_{\textrm{box}} = L_{\textrm{box}}^3$. For the data, we also consider $n_{\textrm{obs}} = N_{\textrm{gal},\textrm{obs}}/V_{\textrm {s}}$, where $V_{\textrm {s}}$ is the comoving volume of the survey. Whilst for the clustering, a measurement of $w_{\textrm{p}}$ is estimated in both the simulation box and the observational data using 13 bins in the projected perpendicular distance range $0.5 < r_{\textrm{p}} / (h^{-1} \textrm{Mpc}) < 50$; this range was selected after testing different choices. For both observational metrics, the uncertainties are estimated using jackknife resampling to account for sample variance, using the full covariance matrix for $w_{\textrm{p}}$ (see, for example, \citealt{Norberg2009}). As we combine these measurements to fit the HOD model to the observational data, we need to make sure that this results in catalogues with accurate and unbiased measurements of $n_{\textrm{gal}}$ and $w_{\rm p}$. For example, by giving the majority of the weight to the clustering by fitting $w_{\textrm{p}}$ only, we would end up with a good reproduction of the measured two-point galaxy statistic, but we would miss the target number density by around 15-20 per cent, as shown by \cite{Parejko2013}. Such a result would have a strong influence on the calculation of the marked correlation function, which would in turn have an impact on the utility of this test to probe modified gravity, by adding systematic uncertainties in the ranges where we expect the models to differ. On the other hand, by giving more weight to the number density and less to the clustering, we will obtain poorer reproductions of the clustering. The range of ``acceptable" HOD parameters will also be broader in the limit of giving increasing weight to the number density, as we are effectively trying to constrain the 5 HOD parameters from, in the limit, one measurement. Hence, a compromise is required in which both observational measurements are recovered without biases \textGreen{or tensions} at an adequate statistical level of confidence. 

We ran the autocorrelation time analysis and the G-R diagnostic to test the convergence of the MCMC chains for three choices of weight values. 
By calculating the value of $\tau_{\textrm{f}}$, we ensure that the chain has been running for a sufficient number of steps. For cases (1) $(A_{\rm n}, A_{w_{\rm p}}) = (0.15,\ 0.85)$ and (2) $(A_{\rm n}, A_{w_{\rm p}}) = (0.5,0.5)$, $\tau_{\textrm{f}} \sim 450$, which is the number of samples needed for the chain to forget where it started. Following the estimated number for the convergence suggested by \texttt{emcee}, these models need at least $20\,000$ iterations. Case (3) with $(A_{\rm n}, A_{w_{\rm p}}) = (0.85, 0.15)$ converges faster with $\tau_{\textrm{f}} \sim 300$, which is expected, as this model allows a wider range of HOD parameters as a result of the smaller weight assigned to the clustering in the metric. We also compute the G-R diagnostic for the total samples in the different chains, obtaining $R = 1.149$, for case (1), $R = 1.087$ for case (2), $R = 1.071$ for case (3). Convergence is assumed to have occurred for values of  $R<1.2$, though a value of $R=1.1$ or more is considered to be on the large side \citep{Rstat}.  Although, all of the weight cases can be considered as having formally converged according to the $R$ values reported above, the higher $R$ value for case (1) disfavours the weighting scheme where $A_{\rm n} = 0.15$ and $\ A_{w_{\rm p}} = 0.85$.

We illustrate the implications of using different weight \textGreen{values} for the estimation of the number density in Fig.~\ref{fig:n_gal}. The three panels show the distribution of the recovered number density values obtained from the HOD parameters sampled, denoted by $n_{\textrm{sim}}$. When we compare the distributions to the value from the observational sample, $n_{\textrm{obs}}$, we can test how good these fits are, paying attention to any systematic shifts. For the first weight case, $A_{\rm n}=0.15$, $A_{w_{\rm p}}=0.85$ (left panel), there is a mismatch between the mean of the distribution of recovered $n_{\textrm{sim}}$ values and the observed value $n_{\textrm{obs}}$. By comparing the distribution of $n_{\textrm{sim}}$ with a Gaussian distribution with the same standard deviation, we find that \textGreen{there is a tension of} around $1$-$\sigma$ \textGreen{(indicated by the blue line and arrows) between the peak of the histogram and the observed value. Although not formally statistically significant this tension means that less than half of the mocks drawn from the red histogram would have a number density that is in close agreement with the observed value}. In comparison, the other weight \textGreen{value cases} (shown in the middle and right panels of Fig.~\ref{fig:n_gal}) yield more accurate estimates of $n_{\textrm{obs}}$. We note that this behaviour is not observed for $w_{\rm p} (r_{\rm p})$, as this is a function with more bins, where the weights do not have a significant effect on the quality of the fit. 

Another argument we can use to help choose the correct weighting scheme is to examine the form of the $\Delta \chi^2$ distribution. In Fig.~\ref{fig:chi2dist} we show this distribution for the two weight cases that yield similar results for the fit to $n_{\textrm{gal}}$, cases (2) and (3). As we use a model with five free parameters (i.e. the HOD model parameters), we expect that the analytic form of the $\chi^2$ distribution with five degrees of freedom will match that recovered from the MCMC chains. In case (3) the higher weight given to $A_{\textrm{n}}$ reduces the effective number of degrees of freedom to $\nu=4$, so the analytic form of the $\chi^2$ distribution with $\nu=4$ is a better match to the histogram of values from the MCMC chains. This is expected as more weight is given to one specific bin, the number density value, rather than the clustering. It is only when we give the same weight to both metrics that the expected value of $\nu = 5$ is recovered. This is relevant to consider when choosing the best weighting scheme as we ensure that the contribution of these two metrics is consistent with the model we have implemented to fit the data. 
After comparing the different panels of Fig.~\ref{fig:n_gal} and the results shown in Fig.~\ref{fig:chi2dist}, we chose a weight $A_{\rm n} \sim 0.5$ to obtain a \textGreen{more accurate} estimate of $n_{\textrm{obs}}$ and $w_{\textrm{p},\textrm{obs}}$. 

Once we fix the weight \textGreen{values} to $A_{\textrm{n}}=0.5,\ A_{w_{\textrm{p}}}=0.5$, we plot the posterior distribution of the parameters in our model in Fig.~\ref{fig:MCMC_posterior}. This plot shows what the parameter space likelihood looks like and how different parameters are correlated. Some of these correlations are expected, like the dependencies between $M_{\rm{min}}$ and $\sigma$ that control the occupation rate of central galaxies in low-mass haloes. Other correlations are more unexpected, like the one between $\sigma$ and $M_0$, where the latter parameter controls the haloes that contain satellite galaxies, once the low-mass haloes with centrals have been fixed. There are no significant differences between the parameter space distributions for the different weighting schemes, apart from slightly wider preferred regions obtained in case (3), which can be inferred from Fig.~\ref{fig:chi2dist}.

\begin{figure}
    \centering
    \includegraphics[width=0.45\textwidth]{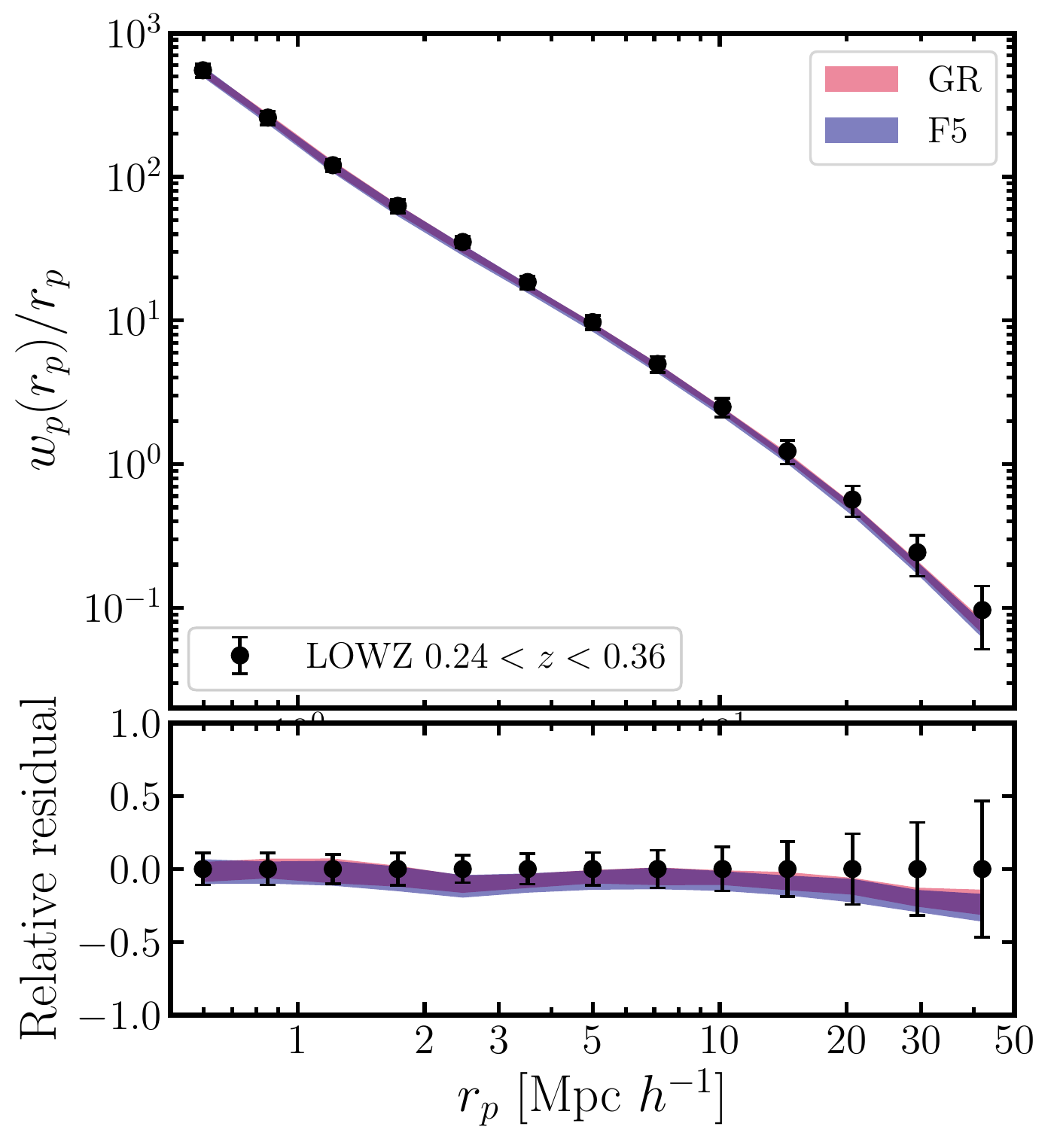}
    \caption{The projected correlation function $w_{\textrm{p}}(r_{\textrm{p}})$ as function of the projected separation, $r_{\textrm{p}}$, for galaxy catalogues created using the HOD samples shown in Fig.~\ref{fig:HODcurves}. The red region corresponds to that covered by all the $w_{\textrm{p}}/r_{\textrm{p}}$ curves, and the black dots show the measurement from the LOWZ sample that we used to fit the model. Uncertainties on the observational measurements have been calculated using jackknife resampling. The bottom subpanel shows the residuals relative to the observational data.}
    \label{fig:2pcf}
\end{figure}

In Fig.~\ref{fig:HODcurves} we show the resulting HOD functions for the galaxy catalogues for both the GR and F5 models, once we fit the model to the observational data. We focus on the example with $A_{\rm n}=0.5$ and $A_{\rm w_p}=0.5$, plotting a random selection of 1000 HOD curves sampled from the acceptable parameter space we find using the MCMC analysis. For the three different \textGreen{weight values} we find similar results in terms of the range of values covered by the HOD parameters. An interesting feature of these HOD parameters is that all three weight cases studied permit $\sigma_{\log M} = 0$, which corresponds to a sharp cutoff in the mass of low-mass haloes that can host a central galaxy. We find that, in general, \textGreen{the weight value cases} where equal or higher weight is given to the clustering, i.e. those with $A_{\rm w_p} = 0.5$ or $ 0.85$, cover the same parameter space. Whereas the model that gives more weight to number density (i.e. the one with  $A_{\rm n}=0.85$) leads to a broader parameter range for those parameters that contribute less to the number density, such as $\sigma_{\log M}$ and $\alpha$, but gives tighter constraints on those that contribute more, such as $M_{\rm{min}}$. We show in Fig.~\ref{fig:2pcf} the results for $w_{\rm p}$ for the same run and models as shown in Fig.~\ref{fig:HODcurves}. In this case, we show the region covered by the individual $w_{\rm p}$ functions selected within the $1$-$\sigma$ region for the HOD parameters, which means that the shaded region represents the uncertainties in the projected correlation function due to the variation in the values of the HOD parameters that are considered as equally good fits. Again, for the three weight cases considered we see the same features, as expected: the clustering is degenerate with the number density for the range of the HOD parameters we find, and the measurement of $w_{\rm p}$ is unbiased for the different weighting schemes, for both the GR and F5 models. These results indicate a good fit to the clustering overall, with a small deviation at large scales, $r_{\rm p} > 20 \,h^{-1} \,\textrm{Mpc}$. Nevertheless, this is smaller than the uncertainties from the jackknife resampling. Additionally, our measurements of $w_{\rm p}$ are also consistent with those from \cite{Parejko2013}, including the small deviation between the mocks and the data at large scales.

\begin{figure*}
    \centering
    \begin{tabular}{cc}
      \vspace*{0.2cm}
      \includegraphics[width=0.44\linewidth]{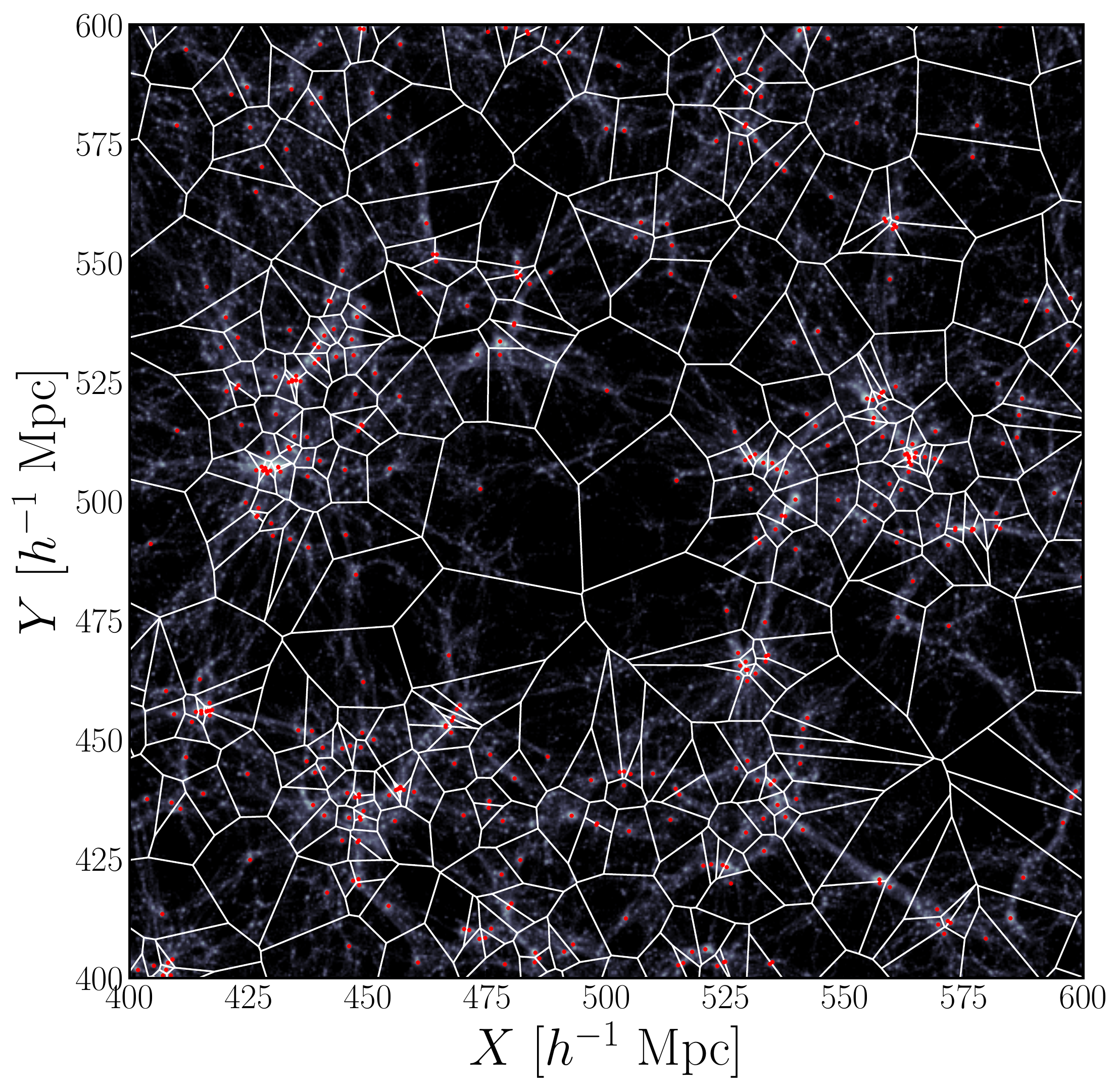}&
      \includegraphics[width=0.495\linewidth]{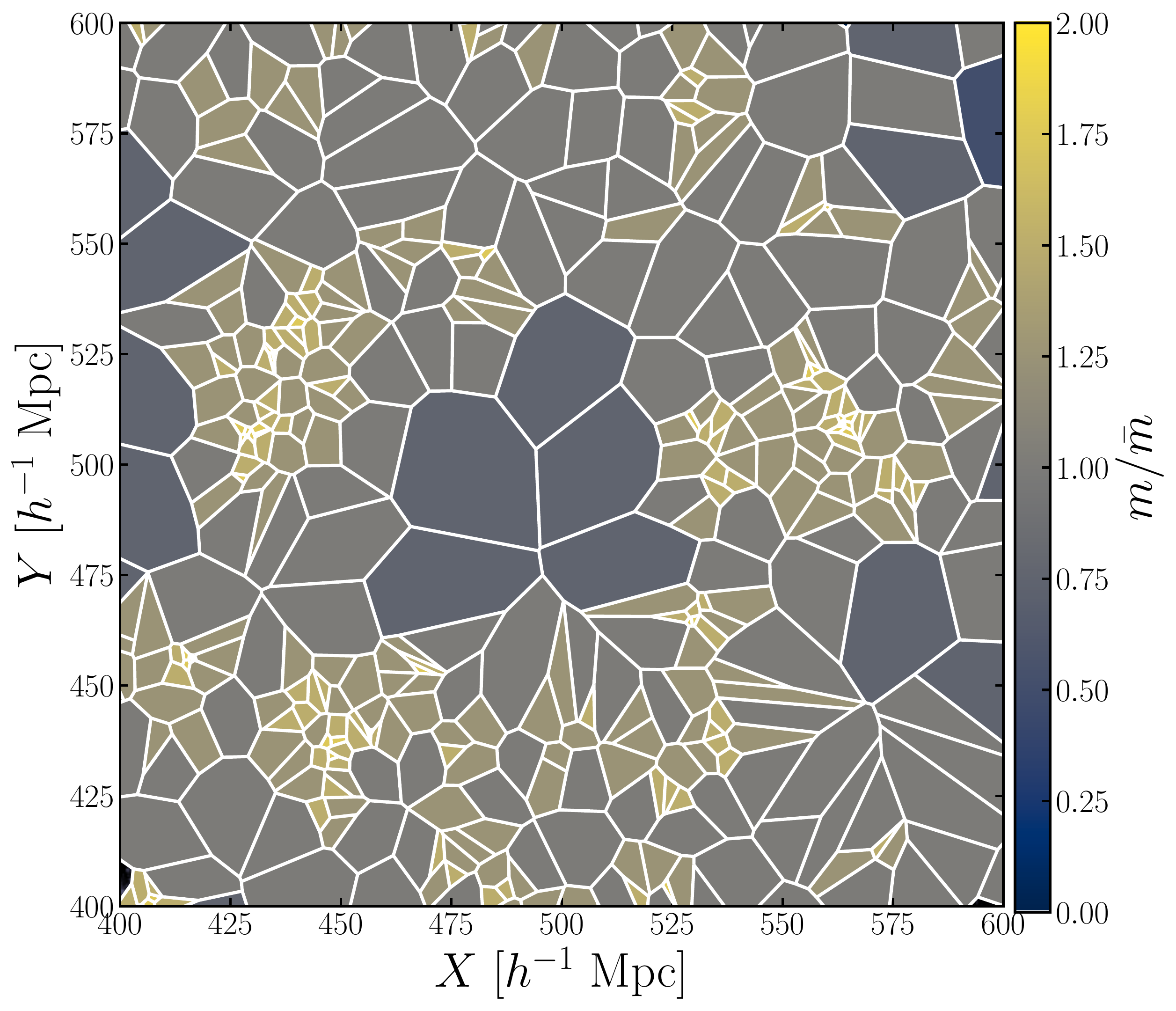} 
    \end{tabular}
    \caption{Left: Voronoi tessellation of the overlying galaxy distribution (red points) for a slice of thickness $\Delta Z = 40\, h^{-1}\, \textrm{Mpc}$ of the GR simulation matter distribution (grey points). The polygons indicated by the white lines are calculated using Voronoi tessellation for the slice projected in the XY plane. The tessellation generates a set of polygons each containing a galaxy, based on the galaxy's nearest neighbours. We use this area to estimate a value of the local density $\rho_i$ for the galaxies in the sample. Right: same as in the left panel but with the individual Voronoi cells coloured according to the value of the mark $m$ of the galaxy in that cell, divided by the mean mark $\bar{m}$. Marks are defined as a mathematical function of $\rho$ (Eqn.~\ref{chp6:eq:mark_rho}) and used in the clustering estimators. Colours indicate by what factor of the mean mark $\bar{m}$ the weight is boosted.}
    \label{fig:Voronoi}
\end{figure*}

\section{Marked correlation function}\label{sec5}

The idea of using the marked correlation function as a new probe of large-scale structure and gravity has been tested using mock galaxy catalogues \citep{Armijo2018,Hernandez2018}, motivated by the theoretical background presented in \cite{White:2016}, who used perturbation theory to explore the properties of the marked correlation function. In these studies, different definitions of the weights applied to galaxies in the marked correlation function were investigated, including ones based on the local density of individual galaxies, the gravitational potential of different environments, and the host halo mass. All of these properties are expected to differ from those in the $\Lambda$CDM paradigm when calculated in modified gravity models, even once the 2-point clustering and abundance have been matched between models. \cite{Satpathy2019} tested the density mark from \cite{White:2016}, applying this to mocks of the LOWZ galaxy sample, using the marked correlation function defined in redshift-space. These authors concluded that their results are limited by the accuracy of the modelling of small scales in the simulations, where most of the differences between GR and MG models were found in previous studies. No significant deviations from $\Lambda$CDM were found by Satpathy et~al. on scales between $6 < s / (h^{-1}\, \textrm{Mpc}) < 69$. The simulations used in \cite{Satpathy2019} have limited resolution, which can affect the results on small scales, which motivates us to refine some aspects of their analysis. Furthermore, the analysis of Satpathy et~al. is in redshift space, which is dominated by the pair-wise velocity distributions on small scales that require further modelling of differences between GR and modified gravity models. Here, we use projected clustering to avoid such complications.

Following \cite{White:2016}, we define the marked correlation function as

\begin{equation}
    \mathcal{M}(r) = \frac{1 + W(r)}{1 + \xi(r)}, \label{chp6:eq:MCF}
\end{equation}
where $\xi(r)$ is the two-point correlation function and $W(r)$ is the weighted or marked version of $\xi$. 
To implement the measurement of the marked correlation function we simply include the marks as additional weights in the correlation function estimator, where the pair counts are replaced by the multiplication of the weights for each galaxy in the pair. We count pairs from the data and random catalogues, redefining the terms in the correlation function estimator to include the mark:
\begin{eqnarray}
    DD & = &\frac{1}{N_g(N_g-1)} \sum_{ij} w_{\textrm{gal},i} w_{\textrm{gal},j},\label{chp6:eq:mDD}\\
    DR & = & \frac{1}{N_gN_r}\sum_{ij} w_{\textrm{gal},i} w_{\textrm{ran},j},\label{chp6:eq:mDR}\\
    RR & = & \frac{1}{N_gN_r}\sum_{ij} w_{\textrm{ran},i} w_{\textrm{ran},j}\label{chp6:eq:mRR},
\end{eqnarray}
where $w_{\textrm {gal,i}}$ is the value of the total weight for each galaxy, and $w_{{\rm ran},i}$ is the counterpart for a random point. This is made up of the weight to compensate for observational effects, such as the radial selection function and the redshift completeness, $w_{{\textrm{obs},i}}$, and the mark to give a total weight of  

\begin{equation}
    w_{ {\rm gal},i} = {m_{i}} w_{ {\rm obs},i},
    \end{equation}

Randoms are marked by the mean mark $\bar{m}$ so that the total weight for a random is 

\begin{equation}
    w_{ {\rm ran},i} = \bar{m} w_{ {\rm obs},i}.
 \label{chp6:eq:w_tot}   
\end{equation}

We use the same prescription employed by  \cite{Satpathy2019} to ensure that the weighted correlation functions depend on the local densities around galaxies. 
For a density-motivated definition, the mark uses an estimation of the local density of an individual galaxy, $\rho_i$, which is defined as the inverse of the volume associated with a galaxy in the density field, in units of the mean density $\bar{\rho}$ of the field. Then we define a density-based mark of the form
\begin{equation}
    m = \left( \frac{\rho}{\bar{\rho}} \right)^p, \label{chp6:eq:mark_rho}
\end{equation}
where $p$ is a free parameter we can vary, to up-weight different density environments. For example, a selection of $p<0$ up-weights low-density regions, where the additional gravity force in MG is prevalent. On the other hand, with $p>0$, high-density environments are favoured, and halos in unscreened regimes can be tested. Note that any normalization of $\rho$ introduced in Eqn.~\ref{chp6:eq:mark_rho} will be included in the value of $\bar{m}$ in the estimators of Eqns.~\ref{chp6:eq:mDD}, \ref{chp6:eq:mDR} and \ref{chp6:eq:mRR}. These definitions produce similar results in distinguishing MG from GR to those obtained using the log-transform density field power spectrum or the clipped density field statistic \citep{Valogiannis2018}. 

Instead of measuring the correlation function in redshift-space, $\xi(s)$, as was done by \cite{White:2016} and \cite{Satpathy2019}, we decide to use $w_{\rm p}(r_{\rm p})/r_{\rm p}$, the projected correlation function divided by the projected pair separation perpendicular to the line of sight, $r_{\rm p}$. This is approximately a real-space quantity. Hence, we avoid dealing with the modelling of redshift-space distortions, which would add a layer of complication (see e.g. \citealt{Cuesta2020,Carol:2023}) and can weaken any conclusions by introducing noise. Currently, RSD modelling performs best on intermediate to large scales, where it is more challenging to distinguish modified gravity from GR \citep{Paillas2018}. Another reason for choosing to work in real space is that the effects of RSD modify the local densities obtained from the Voronoi tessellation, as shown in \cite{Armijo2018}, which reduces the signal of modified gravity in the amplitude of the marked correlation function. Finally, measuring RSD on these scales to test modified gravity is not within the scope of this study, which is already known to be difficult to model for $f(R)$ theories \citep{Hernandez-Aguayo2019}. In the next section, we explain more about the choice and calculation of density-dependent galaxy marks.

\subsection{Local density estimation: the Voronoi tessellation}
We base the estimation of the local galaxy density on Voronoi tessellation \citep{Voronoi1908} in 2D as we are focusing on projected-real space clustering. Voronoi tessellation is a computational method to partition a space according to a given geometrical criterion. The Voronoi tessellation is defined in general by a $n$-plane with $N$ points, where each point generates a $n$-polytope\footnote{The $n$-dimension generalization of a polyhedron.} that contains all of the region closer to that point than to any other. The estimation of the local density for our galaxies is performed in a 2D projection of the original $\textrm{XYZ}$ 3D Cartesian coordinates. For the simulations, this is a straightforward procedure. In our case, a galaxy sample generates a set of Voronoi cells in two dimensions, each with an area, coming from a projected local volume. We choose a thin 3D slice with width $40  (h^{-1}\ \textrm{Mpc})$, which is selected to maximize the number of galaxies projected in each area, whilst at the same time minimising cutting off individual structures in the different volumes. (The choice of slice width is discussed further in Paper II). With the  tessellation area, we define an individual volume $V_i$ for each galaxy, since the remaining dimension is provided by the thickness of the slice, and  define the local projected density: 

\begin{equation}
    \rho_i = \frac{1}{V_i}.\label{chp6:eq:rho}
\end{equation}

\textRed{Note that by projecting galaxy positions in slices along the redshift direction before performing the Voronoi tessellation we are effectively applying a smoothing to the galaxy density field, which depends on the depth of the projected slice.} Estimating the local density using the Voronoi approach is a relatively inexpensive and intuitive method, where galaxies in overdense environments will have small volumes associated with them and hence high densities, and more isolated galaxies will have larger volumes and therefore smaller densities. \textRed{Effectively, this is a reconstruction of the density field using galaxies, which shares features with the underlying matter field \citep{Paranjape2020}}. Voronoi tessellations have been used in a wide range of problems in astrophysics and cosmology, such as the identification of cosmic voids \citep{Platen:2007,Neyrinck:2008} and probing the primordial cosmology and galaxy formation \citep{Paranjape2020}. In Fig.~\ref{fig:Voronoi} we show the Voronoi diagram of the galaxy distribution. In the left panel, we show the shape of the actual Voronoi cells in the 2D projection of the $38.4h^{-1}\, \textrm{Mpc}$ thick slice, which comes from one of the HOD catalogues produced from the cubic box simulations. Here, the cells of different sizes are generated by tracers of the underlying matter field and are representative of the environment in which they reside. In the right panel, we relate these Voronoi cells to the actual marks $m$ defined by Eqn.~\ref{chp6:eq:mark_rho}, with an arbitrary positive value for $p$, divided by the value of the mean mark $\bar{m}$. Then, we colour each Voronoi cell to show how different regions are up or down-weighted when the marked correlation function is computed. For example, small scales dominated by clusters and groups of galaxies are boosted when counting pairs, whereas pairs that include more isolated galaxies yield smaller marks. 

\subsection{Results}

\begin{figure}
    \centering
    \begin{tabular}{c}
         \includegraphics[width=0.95\linewidth]{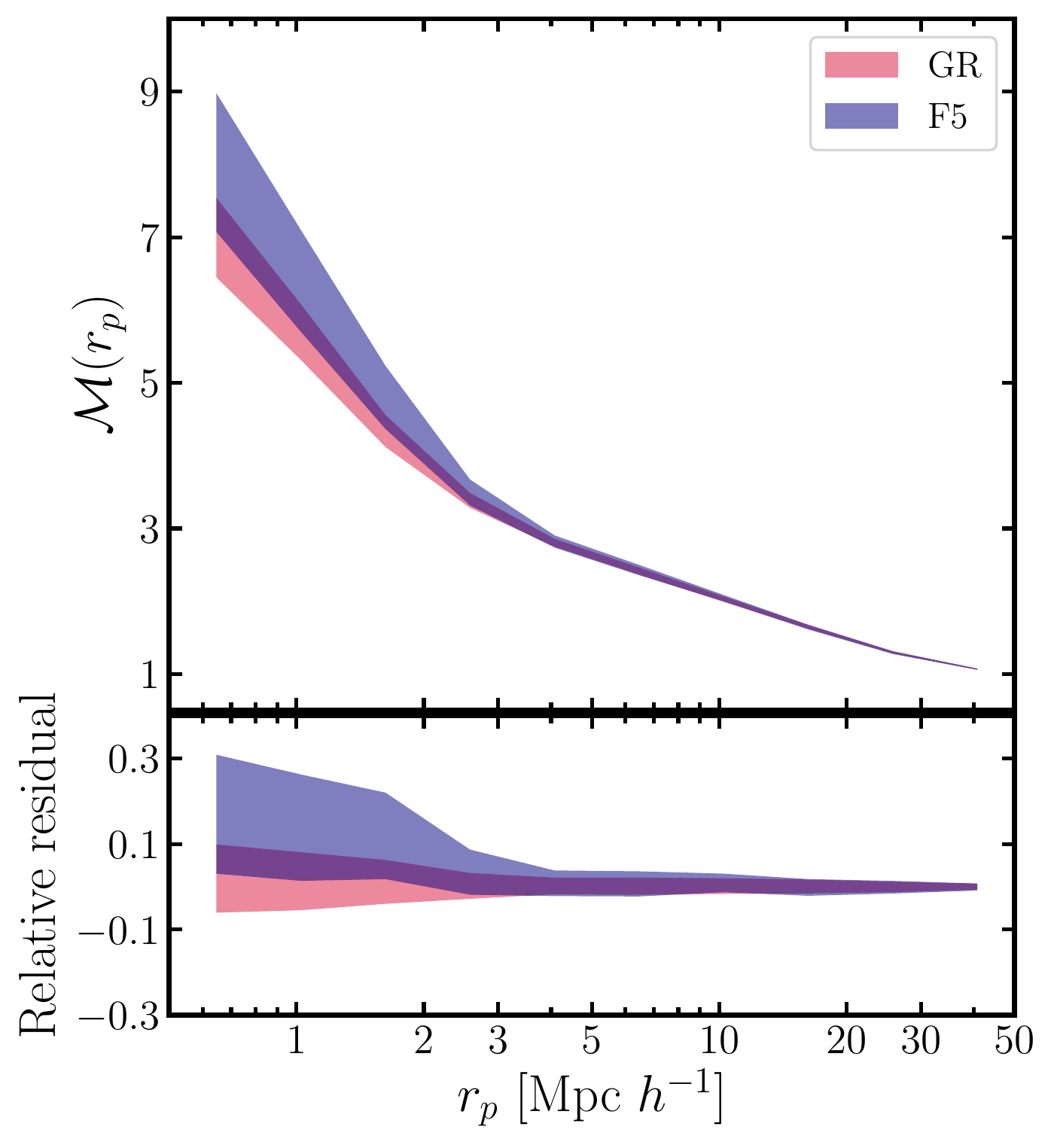}

    \end{tabular}
    \caption{ The marked correlation function $\mathcal{M}(r_p)$ as a function of the projected distance $r_p$ for the HOD mock galaxy catalogues from the GR (red) and F5 (blue) simulations. Top panel: $\mathcal{M}(r_p)$ for the HOD mock catalogues within the $1$-$\sigma$ confidence interval from the MCMC fitting of the two-point clustering and number density of the targeted sample. The shaded areas for the models come from selecting the best-fitting 68\% of HOD catalogues for each model, GR, F5 at redshift $z=0.3$, the mean redshift of the survey, (dark red and dark blue). The bottom panel shows the relative residual taking the median of the GR simulation HOD catalogues as a reference.}
    \label{fig:MCF_LOWZ_CMASS}
\end{figure}

We calculate the marked correlation function for the mock samples using the marks derived from the local density measurements obtained from the Voronoi tessellation. To compute the terms in Eqn.~\ref{chp6:eq:MCF} we use the Landy-Szalay estimator to calculate $\xi(r_{\rm p},\pi)$. When solving the integral in the projected correlation function we consider separations in the line-of-sight direction, $\pi$, using logarithmically spaced bins. By doing this, we achieve better accuracy in the integral calculation for the small $\pi$ separations at which the correlation function changes rapidly. 
We use the publicly available \textsc{twopcf}\footnote{\url{https://github.com/lstothert/two_pcf}} code to compute the $w_{\rm p}(r_{\rm p})$ for the data and mock catalogues; this code supports logarithmic binning and estimators using weighted pairs. The code can also efficiently calculate jackknife errors in a single loop over the galaxy pairs. For the mock catalogues, we select a random sample of 1000 HOD parameter sets selected from the posterior distribution obtained in Section~\ref{sec4}. 
To study the marked statistic of the HOD mock catalogues we select the central 68 per cent of the total sample of values that are closest to the mean of $\mathcal{M}$ for each model. 

We plot the results for the marked correlation functions $\mathcal{M}(r_{\rm p})$ of the HOD mock catalogues in Fig.~\ref{fig:MCF_LOWZ_CMASS}. We compare $\mathcal{M}(r_{\rm p})$ for the GR and F5 models created from the snapshot at redshift $z = 0.3$, using the random sampling of the HOD parameters within the $1$-$\sigma$ confidence interval region. The model predictions overlap at separations larger than $r_{\rm p}>3\,h^{-1}\, \textrm{Mpc}$. However, for separations $r_{\rm p} < 3 \,h^{-1}\, \textrm{Mpc}$ the models start to diverge, with only a modest overlap in the errors. These are the $r_{\rm p}$ separations where there is the potential to find a significant difference between the model predictions but for a survey with a better measurement of the number density of galaxies and the galaxy clustering than the LOWZ sample considered here (see Paper II).

\section{Conclusions}\label{sec6}
We have introduced a new framework to test gravity on different scales using wide-field surveys. We use galaxies as tracers of the matter field to probe the imprint of modified gravity on the cosmic large-scale structure. Such models aim to provide an alternative to the cosmological constant to explain the accelerating cosmic expansion. The viable model we study presents two interesting features: the screening mechanism invoked to hide the modifications where GR is known to be accurate, and the additional fifth force arising from the new degrees of freedom in modified gravity. Then, this fifth force can be detected in regions of high curvature at cosmic scales, where GR still needs to be tested \citep{Zhang2007,Arai:2023}.

From the theoretical side, and to predict the behaviour of the marked correlation function, we prepare mock galaxy catalogues using simulations of a $\Lambda$CDM-GR universe and compare these with mocks from a simulation which uses the $f(R)$ theory of gravity with fifth force amplitude of $\left| f_{R0} \right| = 10^{-5}$ (using the parametrisation of \citealt{HuSawicki2007}). We use the HOD prescription to populate haloes and subhaloes with central and satellite galaxies, from which we extract the best-fitting parameters in terms of the reproduction of the projected correlation function $w_{\rm p}(r_{\rm p})$ and galaxy number density $n_{\textrm{gal}}$. 

We built a \textGreen{phenomenological} $\chi^2$ using the \textGreen{weighted} individual $\chi^2$ from the measurements of $n_{\textrm{gal}}$ and $w_{\rm p}$, and test different \textGreen{weight values}  to investigate any systematic \textGreen{shifts or tensions} in the recovered quantities. We find that both measurements obtain better results if equal weights are given. \textGreen{This approach suggests ranges of weight values to use to avoid biases in the recovered statistics; with current datasets, these differences are marginal and perhaps best described as tensions (see Paper II). Nevertheless, our approach is objective and reproducible.} The final weight choice is based on the definitions of convergence and the individual chains, in addition to the precision with which the measurements can be recovered. In the case of the number density, if too little weight is assigned to its contribution to the overall $\chi^2$, the target value is not recovered with the uncertainties included, which favours models of the $\chi^2$ where equal weight is given to both the number density and clustering. 
Using the $\chi^2$ distribution, we choose a range of HOD parameters within the $1$- $\sigma$ confidence interval to create mocks for both the GR and F5 simulations. \textGreen{Note that the same weight values are used for both gravity models.}

We produce accurate mock catalogues that match the $n_{\textrm{gal}}$ and $w_{\rm p}$ measured from observational samples. We find the HOD parameters that best fit these observational measurements using the MCMC algorithm, which leads to a set of mock catalogues that we use to predict the form of the marked correlation function. 
These mock catalogues incorporate uncertainties from the HOD modelling in the calculation of the marked correlation function, which is in principle larger than sample variance alone. Density-dependent marks are defined using an estimation of the local galaxy density based on Voronoi tessellation. We calculate the marked correlation function for the samples we generate comparing the two models of gravity. 

For the LOWZ sample considered as an example here, we are not able to distinguish modified gravity at the level of $\left| f_{R0} \right| = 10^{-5}$ (F5 model) from GR, when considering the uncertainties introduced by the HOD modelling. \textRed{This is discussed further in Paper II in which we apply the test introduced here to the LOWZ and CMASS samples, and present more information about the analysis of the observational data.}
Then, the importance of this test is to show how the marked correlation function can deal with these uncertainties, and how it can break the degeneracy of the number density and two-clustering in the context of MG models. In the companion paper, we consider the constraints from other current surveys and speculate on the type of survey that would be needed to differentiate F5 gravity from GR.

\section*{Acknowledgements}
The authors would like to thank Yan-Chuan Cai for helpful conversations regarding the project and Christian Arnold for providing the simulation data. This work was supported by the World Premier International Research Center Initiative (WPI), MEXT, Japan. JA acknowledges support from CONICYT
PFCHA/DOCTORADO BECAS CHILE/2018 - 72190634.
PN and CMB are supported by the UK Science and Technology Funding Council (STFC) through ST/T000244/1. NDP acknowledges support from RAICES, a RAICES-Federal, and PICT-2021-I-A-00700 grants from the Ministerio de Ciencia, Tecnología e Innovación, Argentina.  We acknowledge financial support from the European
Union’s Horizon 2020 Research and Innovation programme under the
Marie Sklodowska-Curie grant agreement number 734374 - Project
acronym: LACEGAL. 
This work used the DiRAC@Durham facility managed by the Institute for Computational Cosmology on behalf of the STFC DiRAC HPC Facility (www.dirac.ac.uk). The equipment was funded by BEIS capital funding via STFC capital grants ST/K00042X/1, ST/P002293/1, ST/R002371/1 and ST/S002502/1, Durham University and STFC operations grant ST/R000832/1. DiRAC is part of the National e-Infrastructure.

\section*{Data Availability}
The simulations used on this study are available and accessible on reasonable request. The data products of this work can be shared upon request to the corresponding author.



\bibliographystyle{mnras}
\bibliography{example} 






\bsp	
\label{lastpage}
\end{document}